\documentclass[12pt]{article}
\usepackage{amssymb,amsmath,epsfig}
\allowdisplaybreaks
\begin{document}
\title{\bf Comprehensive Study of Bouncing Cosmological Models in $f(Q,T)$ Theory}
\author{M. Zeeshan Gul \thanks{mzeeshangul.math@gmail.com} , M. Sharif \thanks{msharif.math@pu.edu.pk}~ and~
Shamraiza Shabbir \thanks{shamraizashabbir@gmail.com}\\
Department of Mathematics and Statistics, The University of Lahore,\\
1-KM Defence Road Lahore-54000, Pakistan.}

\date{}
\maketitle

\begin{abstract}
The main objective of this article is to investigate the viability
of bouncing cosmological scenarios using different forms of scale
factors with perfect matter configuration in the framework of
extended symmetric teleparallel theory. This modified proposal is
defined by the function $f(Q,T)$, where $Q$ characterizes
non-metricity and $T$ denotes the trace of energy-momentum tensor.
We investigate the modified field equations of this theory using
different parametric values of the Hubble parameter and
non-metricity to derive viable solutions. These solutions are
relevant in various cosmological bounce models such as
symmetric-bounce, super-bounce, oscillatory-bounce, matter-bounce
and exponential-bounce models. Furthermore, we examine the behavior
of energy density and pressure to analyze the characteristics of
dark energy. A comprehensive analysis is also conducted to explore
the behavior of the equation of state parameter and deceleration
parameter to examine the evolutionary eras of the cosmos. Our
findings show that the $f(Q,T)$ gravity describes the cosmic
expansion in the vicinity of the bouncing point during the early and
late times of cosmic evolution.
\end{abstract}

\section{Introduction}

Einstein's general theory of relativity (EGTR) revolutionized our
understanding of gravity and spacetime, which plays a crucial role
in modern physics. It has been studied extensively through
observations and experiments on the basis of complex shapes and
measurements in space known as Riemannian geometry. Weyl \cite{1}
provided a comprehensive description of gravitational fields and
matter using a more general framework than Riemann's space with the
aim to unify electromagnetic and gravitational forces. The
Levi-Civita connection plays a vital role in the Riemann space by
comparing vector's length \cite{2}. However, Weyl proposed an
alternative connection that does not consider the magnitude of
vectors during parallel transport. To address this limitation, he
introduced a second connection referred to as the length connection
which focuses on adjusting or measuring the conformal factor without
considering the movement of vector's direction. Beyond Riemannian
geometry, non-Riemannian geometries provide more comprehensive
representations of spacetime curvature incorporating torsion
(turning or rotation) and non-metricity (variation from metric
compatibility). Weyl's theory considers non-metricity through
covariance derivative of the metric tensor that is not equal to zero
\cite{3}. Various extended theories of gravity in different context
has been discussed in \cite{3a}-\cite{3j}.

The non-metricity offers a different cosmological model in the
absence of dark energy (DE). The incorporation of non-metricity into
gravitational theory is driven by a range of mathematical and
physical factors. One compelling reason arises from the geometric
explanation of the non-metricity associated with the metric tensor.
The concept of non-metricity involves the alteration in the length
of a vector as it undergoes during parallel transport which offers
the important insights into the geometric characteristics of
spacetime. There is a growing interest among researchers in
investigating the geometry which involves the non-metricity,
particularly the $f(Q,T)$ theory for multiple reasons including its
theoretical consequences, alignment with observational data and its
importance in cosmic scenarios \cite{4}. This theory presents a new
geometric understanding of spacetime by introducing the
non-metricity as a fundamental quantity. This extended theory
incorporates the trace of energy-momentum tensor (EMT) in the
functional action of symmetric teleparallel theory and effectively
accounts the cosmic accelerated expansion. Researchers are
increasingly interested in exploring various aspects of this theory.
Arora et al \cite{5} examined the characteristics of DE in this
theory using different constraints on the model parameters. The same
authors \cite{6} analyzed that this theory offers a novel approach
in understanding the dark sector of the cosmos. The geometry of
compact stars with different considerations in $f(Q)$ and $f(Q,T)$
theory has been studied in \cite{06a}-\cite{06f}.

Singh and Lalke \cite{8} studied the cosmological implications using
hyperbolic solutions in the context of this theory. Their key
findings demonstrated that this modified theory provides a mechanism
for explaining the accelerated expansion of the cosmos. Xu and his
colleagues \cite{4} examined the dynamics of the universe in the
extended symmetric teleparallel theory and compared the obtained
results with $\Lambda$CDM model. Gadbail and his co-authors \cite{9}
uncovered important cosmological insights in this modified framework
and determined how this theory demonstrates the cosmic phenomena.
Narawade et al \cite{10} analyzed the cosmic acceleration through
various cosmographic parameters in the same theory. Bourakadi et al
\cite{11} demonstrated that this theory yields significant
consequences in the formation and evolution of black holes. Shekh
\cite{12} investigated late-time cosmic acceleration through newly
developed scale factor for the FRW model in this theory. The cosmic
acceleration through deceleration parameter in this context has been
studied in \cite{13}. Sharif and Ibrar \cite{14} explored the
reconstruction of a ghost dark energy model in this context.

Bouncing cosmology offers a compelling approach to addressing
solutions for initial singularity issues \cite{15}-\cite{18}. This
concept aims to resolve the challenges associated with the big bang
singularity, which is a significant problem in cosmology. The
fundamental idea behind bouncing cosmology is to propose a model
where the universe does not originate from a singular point (as
theorized by the big bang), but instead undergoes a contraction
followed by a bounce that leads to its current expansion. This model
helps to avoid theoretical complications and infinite values
resulting from singularities, offering a smooth and more accurate
explanation of the cosmic origin and its dynamic properties.
Furthermore, cosmic bounce has been investigated in \cite{19}, which
addresses several issues of the early cosmos such as flatness
problem, horizon problem and initial singularity.

The investigation of bouncing cosmology in different modified
theories has gained significant attention due to its fascinating
characteristics. Ilyas and Rahman \cite{21} explored the FRW model
in $f(R)$ gravity that addresses the big bang singularity through
bouncing cosmology. Shamir \cite{22} examined the viable bouncing
solutions in the framework of $f(G,T)$ gravity, where $G$ is
Gauss-Bonnet invariant. Stability of the closed Einstein universe
and Noether symmetry approach in the modified famework has been
discussed in \cite{22a}-\cite{22c}. Zubair et al \cite{23} found
that the matter-bounce models exhibit stability only for linear
forms of $f(R,T)$ function while the reconstructed solutions show
instability for the power law model. The same authors \cite{24}
investigated bouncing cosmology in the framework of
$f(\mathcal{T},T)$ gravity, ($\mathcal{T}$ represents the torsion)
by examining cosmographic parameters. Ganiou et al \cite{25} used
the reconstruction approach to analyze the specific $f(G)$ gravity
models that describe the critical phases of the cosmic evolution.
Singh et al \cite{26} explored a bouncing scenario in $f(R,T)$
gravity using parametrization of the Hubble parameter. Their results
offer valuable insights into a cosmological model where the universe
undergoes a bounce, highlighting a key features such as singularity
avoidance, phantom divide crossing and extreme dynamics near the
bouncing point. Houndjo et al \cite{28} discussed the bouncing
cosmology in the context of $f(\mathcal{T})$ theory. Sharif et al
\cite{29} investigated the bouncing cosmology in the framework of
$f(Q)$ gravity using a reconstruction approach with perfect matter
configuration.

This paper is organized as follows. Section \textbf{2} outlines the
fundamental formulation of $f(Q,T)$ gravity. A detailed examination
of different types of bouncing solutions are presented in section
\textbf{3}. In order to evaluate the bouncing cosmos, we calculate
the solution to gravitational field equations using a different
parametrization of scale factor. Additionally, we discuss the
graphical behavior of cosmic parameters including scale factor,
Hubble parameter, fluids parameter and EoS parameter. In section
\textbf{4}, we examine the behavior of the deceleration parameter
and analyze the energy conditions, redshift parameter to assess the
viability of a non-singular bounce. Our main findings are summarized
in section \textbf{5}.

\section{$f(Q,T)$ Theory and FRW Universe Model}

The corresponding integral action is defined as \cite{3b}
\begin{equation}\label{1}
S=\frac{1}{2}\int f(Q,T)\sqrt{-g}d^{4}x+\int
\mathcal{L}_{m}\sqrt{-g}d^{4}x.
\end{equation}
The non-metricity is given by
\begin{equation}\label{2}
Q=-g^{\gamma\beta}(L^{\alpha}_{~\sigma\gamma}L^{\sigma}
_{~\beta\alpha}-L^{\alpha}_{~\sigma\alpha}L^{\sigma}
_{~\gamma\beta}),
\end{equation}
where the disformation tensor is defined as
\begin{equation}\label{3}
L^{\varphi}_{~\alpha\zeta}=-\frac{1}{2}g^{\varphi\delta}
\big(\nabla_{\zeta}g_{\alpha\delta}+
\nabla_{\alpha}g_{\delta\zeta}-\nabla_{\delta}g_{\alpha\zeta}\big).
\end{equation}
The superpotential is expressed as
\begin{equation}\label{4}
P^{\alpha}_{~\zeta\lambda}=-\frac{1}{2}L^{\alpha}_{~\zeta\lambda}+
\frac{1}{4}(Q^{\alpha}-\tilde{Q}_{\alpha})g_{\zeta\lambda}
-\frac{1}{4}\delta^{\alpha}~_{(\zeta}Q_{\lambda)}.
\end{equation}
The relation for non-metricity using superpotential is given by
\begin{equation}\label{5}
Q=-Q_{\alpha\zeta\lambda}P^{\alpha\zeta\lambda}=
-\frac{1}{4}\big[-Q^{\alpha\zeta\lambda}Q_{\alpha\zeta\lambda}+2Q
^{\alpha\zeta\lambda}Q_{\lambda\alpha\zeta}-2 Q^{\varphi}\tilde {Q}
_{\varphi}+Q^{\varphi}Q_{\varphi}\big].
\end{equation}
The corresponding field equations are
\begin{eqnarray}\nonumber
T_{\alpha\beta}&=&-\frac{1}{2}fg_{\alpha\beta}-\frac{2}{\sqrt
-g}\nabla^{\zeta}(f_{Q}\sqrt-g P_{\zeta\alpha\beta})
-f_{Q}(P_{\zeta\alpha\lambda}Q^{~~\zeta\lambda}_{\beta}
\\\label{6}
&-&2 Q^{\zeta\lambda}_{\quad\alpha}P_{\zeta\lambda\beta})
+f_{T}(T_{\alpha\beta}+\theta_{\alpha\beta}).
\end{eqnarray}
Here, $f_{Q}$ and $f_{T}$ represent the derivatives corresponding to
non-metricity and trace of EMT, respectively.

We consider a flat FRW universe model with scale factor $a(t)$ as
\begin{equation}\label{7}
ds^{2}=dt^{2}-a^{2}(t)(dx^2+dy^2+dz^2).
\end{equation}
The isotropic matter configuration is given by
\begin{equation}\label{8}
T_{\alpha\beta}=(\rho+p)u_{\alpha}u_{\beta}-p g_{\alpha\beta},
\end{equation}
where $\rho$, $p$ and $u_{\alpha}$ represent the energy density,
pressure and four-velocity of the fluid, respectively. Using
Eqs.\eqref{6}-\eqref{8}, the resulting field equations are
\begin{eqnarray}\label{9}
\rho&=&-\frac{1}{2}f-6H^{2}f_{Q}-f_{T}(\rho+p),
\\\label{10}
p&=&\frac{1}{2}f+2f_{Q}\dot{H} +2Hf_{QQ}+6H^{2}f_{Q},
\end{eqnarray}
with
\begin{equation}\label{11}
Q=-6H^{2}, \quad T=\rho-3p.
\end{equation}
Here, $H=\frac{\dot{a}}{a}$ is the Hubble parameter and dot
demonstrates the derivative with respect to time. These field
equations are in complex form due to the involvement of multivariate
functions and their derivatives. To address this challenge, we
consider a specific model as
\begin{equation}\label{11a}
f(Q,T)=\xi_{1}Q+\xi_{2}T,
\end{equation}
to simplify the field equations and obtain explicit expressions for
energy density and pressure. Here, $\xi_{1}$ and $\xi_{2}$ are
non-zero arbitrary constants. Numerous studies have been conducted
on this model in the literature \cite{1z5b}. This model considers a
linear relationship between non-metricity and trace of EMT, helping
us to understand the gravitational phenomena and allows for more
accurate solutions. Consequently, it is regarded as a valuable
theoretical framework to comprehend the fundamental principles of
gravitational physics and carries importance for both theoretical
analysis and practical applications. The corresponding field
equations are
\begin{eqnarray}\label{11b}
\rho&=&\frac{1}{6\xi_{2}}[4\xi_{1}\xi_{2}\dot{H}+6\xi_{1}H^{2}
-18\xi_{1}\xi_{2}H^{2}],
\\\label{11c}
p&=&\frac{1}{9\xi_{2}}[12\xi_{1}\xi_{2}(4\xi_{1}
\dot{H}+6\xi_{1}H^{2})-18\xi_{1}\xi_{2}^{2}H^{2}
+4\xi_{1}\xi_{2}^{2}\dot{H}+6\xi_{1}\xi_{2}H^{2}].
\end{eqnarray}
In the following sections, we explore the behavior of various
bouncing models, providing valuable insights into the structure of
cosmic evolution.

\section{Bouncing Models}

This section examines the viability of various bouncing models like
symmetric-bounce, super-bounce, oscillatory-bounce, matter-bounce
and exponential-bounce II due to their intriguing properties. This
approach enables us to determine the gravitational model based on a
selected cosmological framework which can be derived using different
forms of scale factors and Hubble parameters \cite{1z1}. To obtain a
comprehensive analysis, it is crucial that the above bounce models
must represent the dynamical behavior throughout various cosmic
eras. This can be achieved by analyzing different ranges of the
parametric values to reconstruct different cosmic epochs \cite{1z2}.
Different types of bouncing model are outlined below.

\subsection{Evolution of Symmetric-Bounce Model}

This model was first examined by Cai et al \cite{1z3} to formulate
the non-singular bounce after an ekpyrotic contraction phase. The
idea of a symmetric-bounce is based on the notion that the cosmos
undergoes a phase of contraction from the previous state of
expansion, reaching a minimum size (bounce point) and then starts to
expand again. The term symmetric in this context refers to the
behavior of the cosmological dynamics during the contraction and
expansion phases. This concept is significant in theoretical
cosmology, bouncing universe scenario and other modified gravity
theories. These models attempt to address cosmological challenges
such as the nature of the big bang singularity and the origin of the
cosmos \cite{1z4}. We consider the extended symmetric bouncing
cosmology characterized by scale factor as \cite{1z5}
\begin{equation}\label{12}
a=\mathbb{A}~e^{\frac{\eta t^{2}}{t_{\star}^{2}}},
\end{equation}
where $t_{\star}$ and $t$ are an arbitrary and cosmic time,
$\mathbb{A}$ and $\eta$ are positive constants. However, the cosmic
time is measured in gigayears (Gyr). Understanding the evolution of
the scale factor is essential to comprehend how the universe
expands, contracts or undergoes a bouncing phase. The scale factor
is a positive function that quantifies the change in size of the
cosmos, representing its dynamics over time. The graphical behavior
of scale factor is shown in the left plot of Figure \textbf{1} which
shows a positive symmetric pattern, indicating that the scale factor
decreases and increases in a balanced manner on either side of the
bouncing point.
\begin{figure}
\epsfig{file=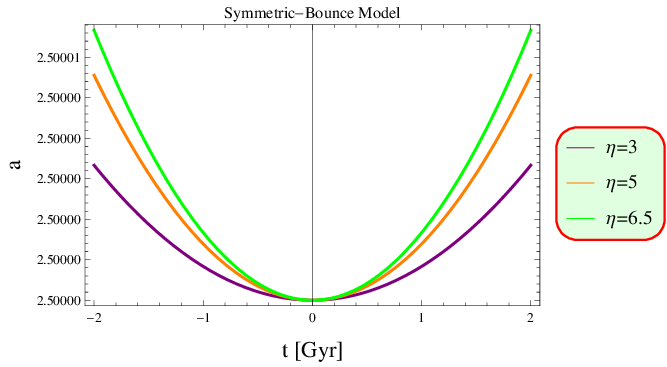,width=.5\linewidth}
\epsfig{file=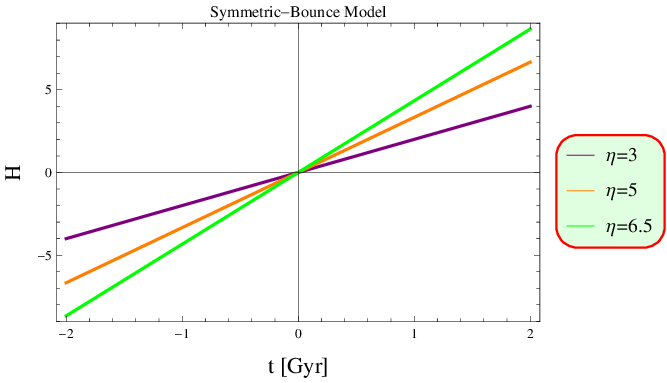,width=.5\linewidth}\caption{Behavior of scale
factor and Hubble parameter versus cosmic time.}
\end{figure}

Using Eq.(\ref{12}), the Hubble parameter and non-metricity become
\begin{equation}\label{13}
H=\big(\frac{2\eta t}{t_{\star}}^{2}\big), \quad
Q=-\frac{24\eta^{2}t^{2}}{t_{\star}^{2}}.
\end{equation}
The right plot of Figure \textbf{1} demonstrates that the Hubble
parameter is zero at bouncing point $(t=0)$ as well as shows
contraction before the bounce $(t<0)$ and expansion after the bounce
$(t>0)$. Using Eq.(\ref{13}) into Eqs.(\ref{11b})-(\ref{11c}), we
obtain
\begin{eqnarray}\label{15}
\rho&=&\frac{4t\eta\xi_{1}\big(-3t(2t^{2}+1)\eta(\xi_{2}-7)+
t_{\star}\xi_{2}\big)}{t_{\star}(2\xi_{2}^{2}-49)},
\\\label{16}
p&=&\frac{1}{t_{\star}^2(2\xi_{2}^{2}-49)}\big[4t\eta\xi_{1}\big(3t(2
t^{2}+1)\eta(\xi_{2}+7)+t_{\star}(3\xi_{2}+14)\big)\big].
\end{eqnarray}
The units for energy density and pressure are considered as $GeV/
cm^{3}$. Figure \textbf{2} shows the variation in density and
pressure for symmetric-bounce model. The graphical behavior
demonstrates that the energy density exhibits an positively
increasing trend and pressure displays negatively downward
trajectory over time. This inverse relationship between energy
density and pressure is in accordance with the expected behavior
predicted by the DE model.
\begin{figure}
\epsfig{file=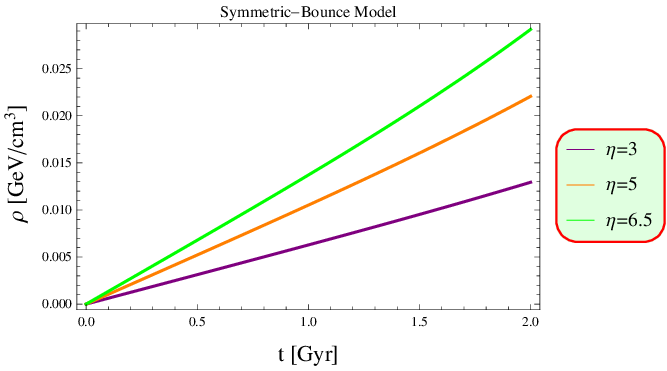,width=.5\linewidth}
\epsfig{file=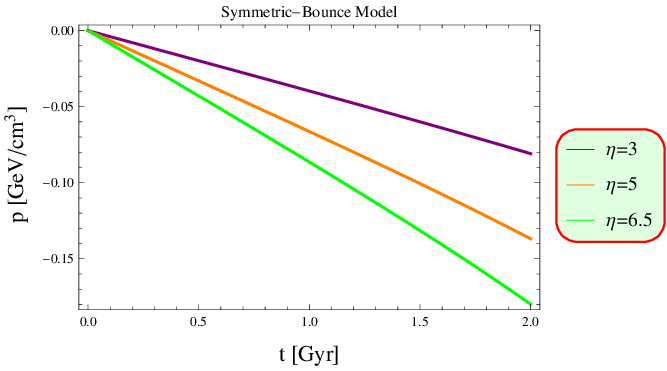,width=.5\linewidth}\caption{Behavior of density
and pressure corresponding to cosmic time.}
\end{figure}

The EoS parameter $(\omega=\frac{p}{\rho})$ can be classified based
on different stages of cosmic evolution. One can obtain
matter-dominated eras such as dust, radiative fluid and stiff matter
for $\omega=0,~\frac{1}{3},~1$, respectively, whereas, the vacuum,
phantom and quintessence phases of the cosmos are characterized by
$\omega= -1,~\omega <-1,~-1<\omega<-\frac{1}{3}$, respectively
\cite{1z5c}. Using Eqs.(\ref{15}) and (\ref{16}), we calculate the
value of the EoS parameter for symmetric-bounce model as
\begin{equation}\label{16a}
\omega=-\frac{3\eta(\xi_{2}+7)\big(2t^3+t\big)+(3\xi_{2}+14)
t_{\star}}{3\eta(\xi_{2}-7)t \big(2t^2+1\big)-\xi_{2}t_{\star}}.
\end{equation}
Figure \textbf{3} shows that the EoS parameter becomes singular at
the bounce point and undergoes rapid evolution in the vicinity of
the bounce. During this period, the EoS parameter exhibits symmetry
around the epoch of the bounce and transitions into the phantom
region $(\omega<-1)$. This behavior indicates a significant shift in
the characteristics of this parameter as the system approaches and
moves towards the bouncing point. In the context of cosmology, such
a bounce represents a critical phase where the cosmos shows
transition from a contracting state to an expanding one.
\begin{figure}\center
\epsfig{file=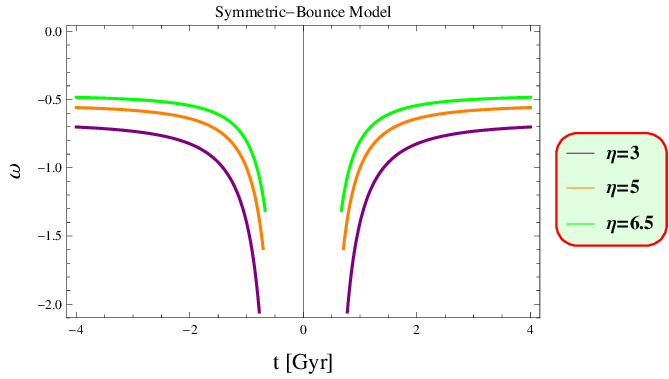,width=.5\linewidth}\caption{Behavior of EoS
parameter for different parametric values.}
\end{figure}

\subsection{Analysis of Super-Bounce Model}

The concept of super-bounce is characterized by a power-law scale
factor which was first proposed in \cite{1z6}. The idea of a
super-bounce suggests that the cosmos undergoes cycles of expansion
and contraction, rather than a single expansion followed by infinite
expansion without encountering a singularity. The general form of
the scale factor is defined as
\begin{equation}\label{17}
a=e^{\big(\frac{t_{b}-t}{t_{0}}\big)^{\frac{2}{\mathbb{Z}^{2}}}},
\end{equation}
where $\mathbb{Z}>3$, $t_{0}>0$ and $t_{b}$ denotes the time of the
bounce event. The graphical behavior of the super-bounce scale
factor is shown in the left plot of Figure \textbf{4}. The graph
indicates that the scale factor is positive, but does not show a
symmetric pattern in this model on either side of the bouncing
point. The corresponding value of the Hubble parameter and
non-metricity become
\begin{equation}\label{18}
H=-\frac{2}{\mathbb{Z}^{2}}\bigg(\frac{1}{t_{b}-t}\bigg),
~~Q=\frac{24}{\mathbb{Z}^{4}}\bigg(\frac{1}{t_{b}-t}\bigg)^{2}.
\end{equation}
In the right plot of Figure \textbf{4} demonstrates the behavior of
the Hubble parameter in super-bounce with distinct characteristics
before and after the bouncing point. This parameter changes its
signatures during the transition phase of contraction/ expansion and
becomes singular at $H=0$. This indicates a critical phase, where
the dynamics of the cosmos undergoes a significant transformation.
Using Eq.(\ref{18}) in Eqs.(\ref{11b})-(\ref{11c}), we have
\begin{eqnarray}\label{19}
\rho&=&-\frac{1}{\mathbb{Z}^{4}(t-t_{b})^{2}(2\xi_{2}^{2}-49)}
\big[2\xi_{1}\big(\big(\mathbb{Z}^{2}+18\big)\xi_{2}-126\big)\big],
\\\nonumber
p&=-&\frac{1}{\mathbb{Z}^{4}(t-t_{b})^{2}(2\xi_{2}^{2}-49)(3\xi_{2}-14)}
\big[36\xi_{1}\big((\xi_{2}(5\xi_{2}-7)+98)
\\\label{20}
&+&7\mathbb{Z}^{2}(\xi_{2}^{2}-28)\big)\big].
\end{eqnarray}
\begin{figure}
\epsfig{file=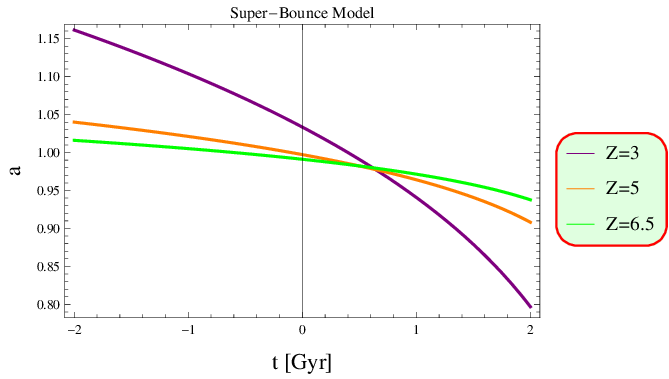,width=.5\linewidth}
\epsfig{file=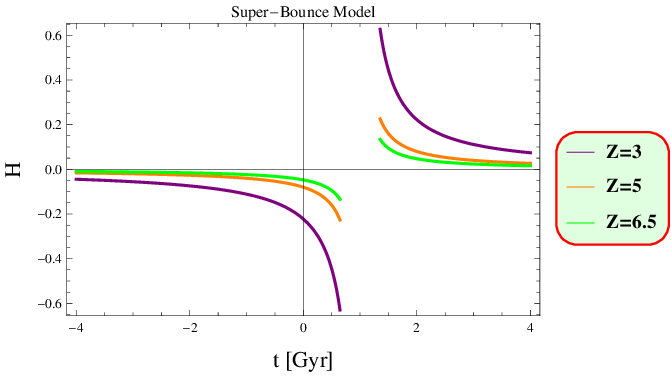,width=.5\linewidth}\caption{Behavior of scale
factor and Hubble parameter versus cosmic time.}
\end{figure}
\begin{figure}
\epsfig{file=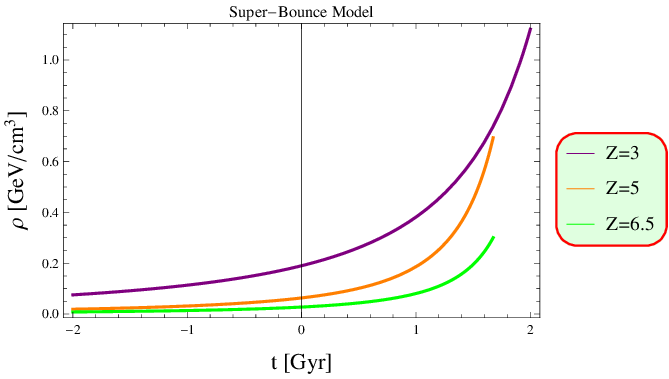,width=.5\linewidth}
\epsfig{file=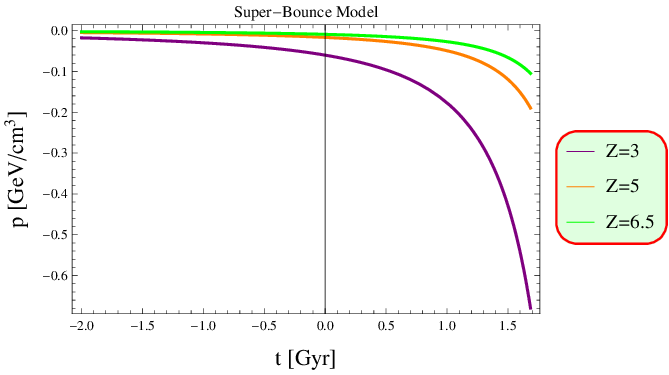,width=.5\linewidth}\caption{Behavior of matter
components versus cosmic time.}
\end{figure}

Figure \textbf{5} depicts the graphical representation of fluid
parameters for the super-bounce model, which are closely resemble
those obtained in the symmetric-bounce model and are consistent with
behavior of the DE model. We use Eqs.(\ref{19}) and (\ref{20}) to
determine the EoS parameter value for super-bounce model as
\begin{equation}\label{20a}
\omega=\frac{1}{(3\xi_{2}-14)\big(\xi_{2}
(\mathbb{Z}^{2}+18)-126)}\big[18\xi_{2}(7-5\xi_{2})+98)+7\mathbb{Z}^{2}
(\xi_{2}^{2}-28)\big].
\end{equation}
Figure \textbf{6} shows that the behavior of EoS parameter is
identical as obtain in symmetric-bounce model. This means that the
physical conditions governing the cosmic behavior is stable and does
not reach infinite values at any point during the bounce which
ensures a smooth transition through this critical phase.
\begin{figure}\center
\epsfig{file=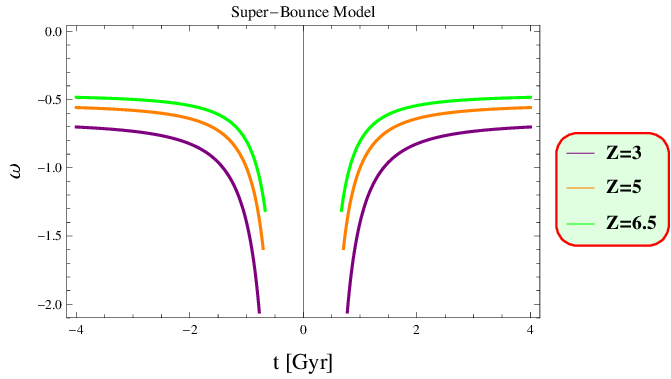,width=.5\linewidth}\caption{Behavior of EoS
parameter corresponding to cosmic time.}
\end{figure}

\subsection{Evolution of Oscillatory-Bounce Model}

The concept of an oscillatory-bounce is significant in the field of
cosmology and scenarios involving the expansion and contraction of
the cosmos. In this bouncing scenario, the cosmos undergoes a
periodic cycles of expansion and contraction. Each cycle initiates
with a big bang followed by a phase of expansion and concludes with
a big crunch, where the cosmos contracts back to a dense state
before beginning with another big bang. This cyclic pattern implies
a repetitive sequence of cosmic events alternating between expansion
and contraction over successive cycles \cite{1z7}. The corresponding
expression for the scale factor is expressed as
\begin{equation}\label{21}
a=\mathbb{A}\sin^{2}\bigg(\frac{\mathbb{B}t}{t_{\star}}\bigg),
\end{equation}
where $\mathbb{A}$ and $\mathbb{B}$ are non-negative constants. In
the left side of Figure \textbf{7} shows the graphical behavior of
scale factor in oscillatory-bounce model. The oscillatory-bounce
model indicates that the cosmos undergoes a periodic cycles of
contraction and expansion. This model shows that there are two
distinct types of bounce events. The first type occurs when
$t=\frac{n\pi t_{\star}}{\mathbb{B}}$, where $n$ is an integer. This
scenario corresponds to a big bang singularity marking a point where
the cosmos contracts to a singular point before expanding again. The
second type of bounce takes place when $t=\frac{(2n+1)\pi
t_{\star}}{2\mathbb{B}}$. This occurs when the cosmos reaches its
maximum size as expansion ends and the cosmos begins to contract
again. This transitional behavior of cosmos is critical in
understanding the dynamics of its evolution as predicted by this
model.
\begin{figure}
\epsfig{file=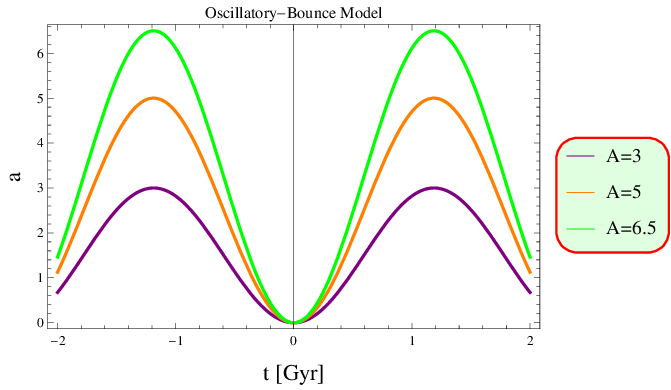,width=.5\linewidth}
\epsfig{file=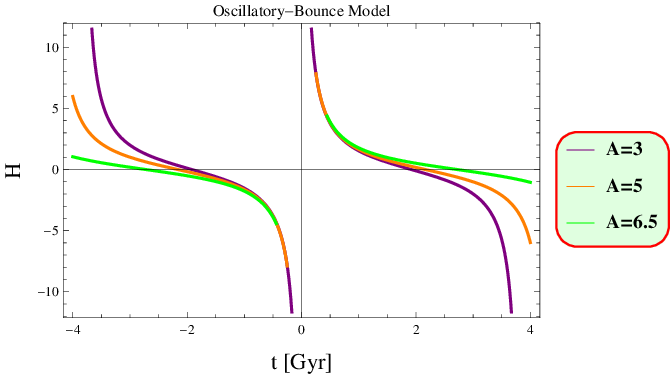,width=.5\linewidth}\caption{Behavior of scale
factor and Hubble parameter versus cosmic time.}
\end{figure}

The expressions for $H$ and $Q$ corresponding to this bouncing model
turn out to be
\begin{equation}\label{22}
H=\frac{2\mathbb{B}}{t_{\star}}\cot\bigg(\frac{\mathbb{B}
t}{t_{\star}}\bigg),
~~Q=-\frac{24\mathbb{B}^{2}}{t_{\star}^{2}}\cot^{2}\bigg(\frac{\mathbb{B}
t}{t_{\star}}\bigg).
\end{equation}
The graphical behavior of the Hubble parameter in the
oscillatory-bounce is shown in the right side of Figure \textbf{7}.
The Hubble parameter becomes singular at bouncing point for
$t=\frac{n\pi t_{\star}}{\mathbb{B}}$. Moreover, the Hubble
parameter undergoes a transition phase of contraction and expansion
at $t=\frac{(2n+1)\pi t_{\star}}{2\mathbb{B}}$. Specifically, it
shifts from positive values to negative values at these points. This
indicates that there is a critical moment, where the Hubble
parameter crosses zero by changing its phase from expansion towards
contraction.

Using Eq.(\ref{22}) in Eqs.(\ref{11b})-(\ref{11c}), we obtain the
field equations corresponding to this model as
\begin{eqnarray}\nonumber
\rho&=&\frac{1}{t_{\star}^{2}(2\xi_{2}^{2}-49)}\big[2
\mathbb{B}^{2}\xi_{1}(-6(\xi_{2}-7)\big(\cot^{2}\bigg(\frac{\mathbb{B}t}
{t_{\star}}\bigg)-2)+(84-13\xi_{2})\\\label{23}
&\times&\csc^{2}\bigg(\frac{\mathbb{B}t}{t_{\star}}\bigg))\big],\\\nonumber
p&=&\frac{1}{t_{\star}^{2}(2\xi_{2}^{2}-49)}\big[2
\mathbb{B}^{2}\xi_{1}(-6(\xi_{2}-7)\big(\cot^{2}\bigg(\frac{\mathbb{B}t}{t_{\star}}\bigg)-2)
+(70-9\xi_{2})\\\label{24}
&\times&\csc^{2}\bigg(\frac{\mathbb{B}t}{t_{\star}}\big))\bigg].
\end{eqnarray}
Figure \textbf{8} depicts the change in matter variables for
oscillatory-bounce model. This model demonstrates an oscillation in
the behavior of the fluid parameters. Prior to the bounce, there is
a positive increase in energy density and it displays a positive
decrease after the bounce. Similarly, pressure exhibits a negative
pattern in the behavior. These graphical behavior support the
current cosmic expansion. Using Eqs.(\ref{23}) and (\ref{24}), we
obtain the corresponding EoS parameter as
\begin{equation}\label{24a}
\omega=-\frac{6 (\xi_{2}+7)\left(\cot^2\left(\frac{\mathbb{B}
t}{t_{\star}}\right)-2\right)+(9\xi_{2}+70)\csc
^2\left(\frac{\mathbb{B} t}{t_{\star}}\right)}{6(\text{$\xi $2}-7)
\left(\cot^2\left(\frac{\mathbb{B} t}{t_{\star}}\right)-2\right)+(13
\text{$\xi $2}-84)\csc^2\left(\frac{\mathbb{B} t}{t_{\star}}\right)}
\end{equation}
Figure \textbf{9} shows that the EoS parameter oscillates over time
reflecting the dynamic nature of the cosmological evolution under
this framework. The graphical representation provides a powerful
insight into the underlying physical processes driving the
oscillations and their impact on the behavior of model.
\begin{figure}
\epsfig{file=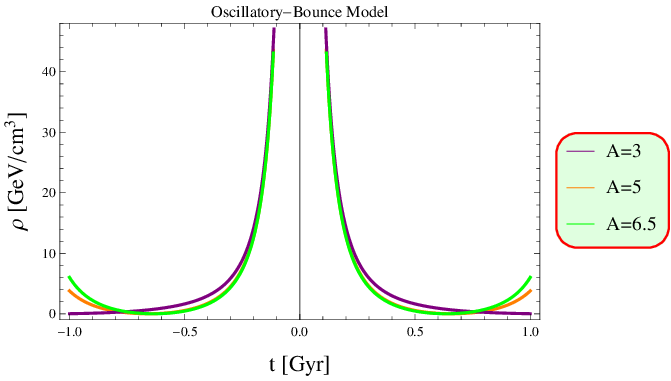,width=.5\linewidth}
\epsfig{file=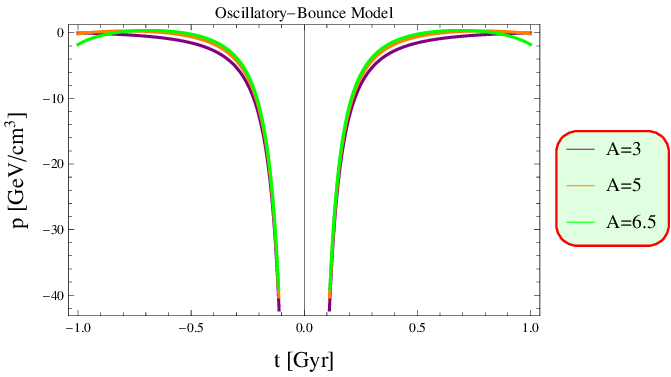,width=.5\linewidth}\caption{Behavior of energy
density and pressure for different values of $\mathbb{A}$.}
\end{figure}
\begin{figure}\center
\epsfig{file=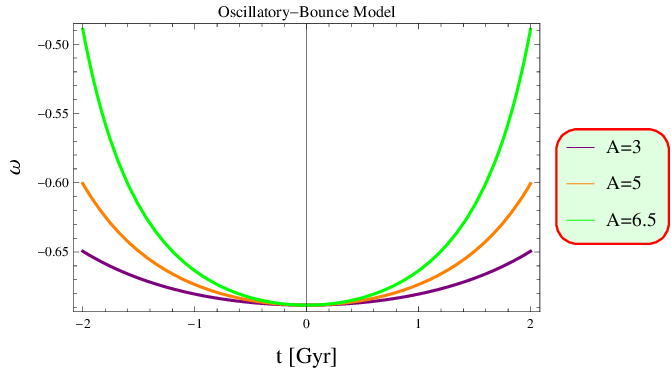,width=.5\linewidth}\caption{Behavior of EoS
parameter for different values of $\mathbb{A}$.}
\end{figure}

\subsection{Study of Matter-Bounce Model}

The matter-bounce scenario is a cosmological model proposed as an
alternative to the big bang theory. In the matter-bounce model, the
cosmos undergoes a phase of contraction before expanding again,
rather than beginning from a singularity as in the big bang theory.
The motivation for the matter-bounce model arises from addressing
some of the theoretical issues present in the standard big bang
model. The matter-bounce model attempts to provide an alternative
framework that avoids initial singularities. An intriguing method
alternative to inflation is the concept of matter-bounce model
\cite{1z9}, which is notable for its compatibility with
observational evidence from the Planck observational data
\cite{1z10}. The corresponding scale factor is expressed as
\begin{equation}\label{25}
a=\mathbb{A}\bigg(\frac{3}{2}\rho_{c}t^{2}+1\bigg)^{\frac{1}{3}}.
\end{equation}
In the given context, $0<\rho_{c}<1$ represents a critical density.
The critical density is a fundamental concept in cosmology used to
understand the fate and geometry of the cosmos based on its overall
density. In the left plot of Figure \textbf{10} demonstrates the
behavior of matter-bounce scale factor is positive and symmetric on
either side of the bouncing point. The Hubble parameter and
non-metricity scalar for this case are given by
\begin{equation}\label{26}
H=\frac{2 t \rho_{c}}{2+3 \rho_{c}t^{2}}, ~~Q=-6\bigg(\frac{2 t
\rho_{c}}{2+3 \rho_{c}t^{2}}\bigg)^{2}.
\end{equation}
\begin{figure}
\epsfig{file=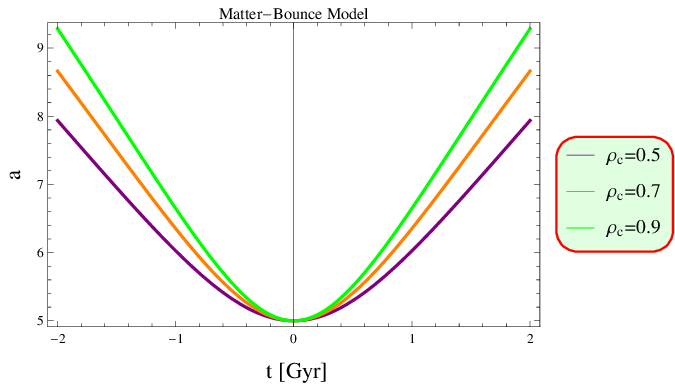,width=.5\linewidth}
\epsfig{file=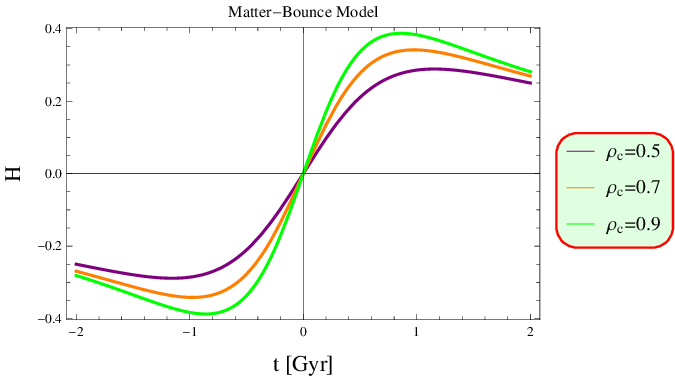,width=.5\linewidth}\caption{Behavior of scale
factor and Hubble parameter to cosmic time.}
\end{figure}

In the right side of Figure \textbf{10} illustrates the behavior of
the Hubble parameter across different phases of the cosmic bounce.
In the pre-bounce phase, the Hubble parameter is negative,
indicating a contracting universe. As the cosmos approaches towards
the critical bounce point, the Hubble parameter reaches to zero
which signifies a momentary end in the contraction and marking the
transition between contraction and expansion phases. By following
the bouncing point during the post-bounce epoch, the Hubble
parameter becomes positive reflecting the cosmic expansion. This
transition through the phases highlight the dynamic nature of the
cosmological model. Substituting Eq.(\ref{26}) into
Eqs.(\ref{11b})-(\ref{11c}), the resulting field equations are
expressed as
\begin{figure}
\epsfig{file=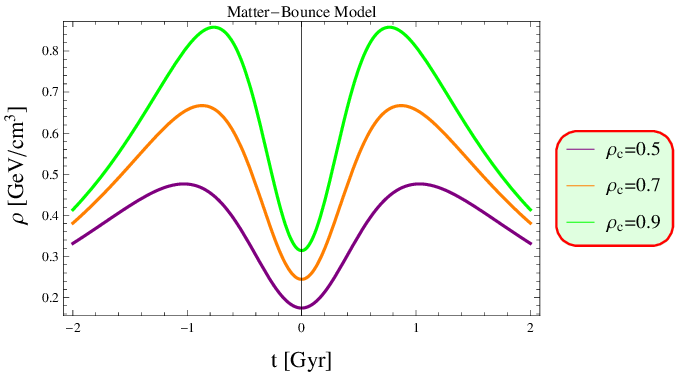,width=.5\linewidth}
\epsfig{file=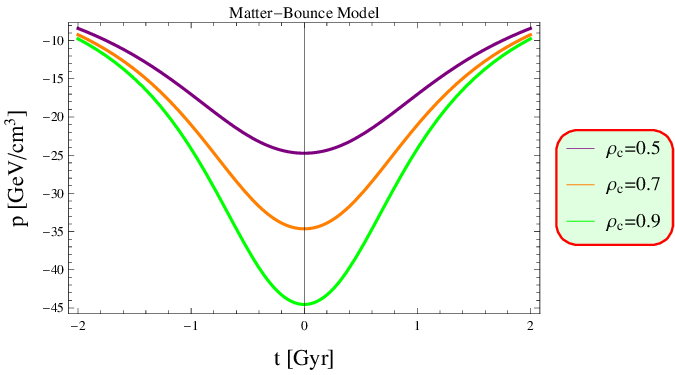,width=.5\linewidth}\caption{Behavior of energy
density and pressure for different values of $\rho_{c}$.}
\end{figure}
\begin{eqnarray}\label{27}
\rho&=&\frac{1}{(2\xi_{2}\rho_{c}^{2}-49)(3
t^{2}\rho_{c}+2)^{2}}\big[2\xi_{1}\rho_{c}(2\xi_{2}-21t^{2}
\rho_{c}(\xi_{2}-6))\big],
\\\label{28}
p&=&\frac{1}{(2\xi_{2}\rho_{c}^{2}-49)(3
t^{2}\rho_{c}+2)^{2}}\big[2\xi_{1}\rho_{c}(28+84t^{2}
\rho_{c}+\xi_{2}(9t^{2}\rho_{c}+6))\big].
\end{eqnarray}
Figure \textbf{11} determines that the behavior of fluid parameters
is consistent with the expected behavior of DE model. Using
Eqs.(\ref{27}) and (\ref{28}), the EoS parameter for this bouncing
model is as follows
\begin{equation}\label{28a}
\omega=-\frac{28+6\xi_{2}+3t^{2}(28+3\xi_{2})\rho_{c}}{-2\xi_{2}
+21t^{2}(\xi_{2}-6)\rho_{c}}.
\end{equation}
Figure \textbf{12} demonstrates that the EoS parameter becomes
singular at the bouncing point and undergoes rapid changes near the
bounce. Notably, the EoS parameter shows symmetry around the
bouncing epoch and exhibits significant evolution in the phantom
region. This evolution describes the dynamical change in the nature
of matter and energy in the cosmos during this critical phase.
\begin{figure}\center
\epsfig{file=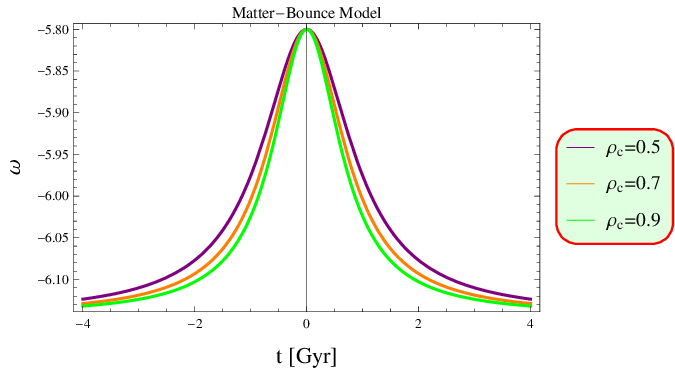,width=.5\linewidth}\caption{Behavior of EoS
parameter for different values of $\rho_{c}$.}
\end{figure}

\subsection{Discussion on Exponential-Bounce Model II}

The exponential model describes the expansion history of the cosmos.
This model is an extension of the original exponential model (also
known as the exponential inflationary universe or exponential
expansion) to explain the rapid expansion of the universe in the
early moments of the big bang. In the context of theories related to
inflation and the dynamics of the cosmic expansion, this model is
referred to a bouncing scenario where the scale factor evolves
exponentially with time. In this bouncing model, the scale factor is
represented as
\begin{equation}\label{29}
a=\mathbb{A}\bigg(\frac{h_{0}}{\varsigma+1}(t-t_{b})^{\varsigma+1}\bigg),
\end{equation}
where $h_{0}$ and $\varsigma$ are an arbitrary constant. The
graphical representation of scale factor is shown in the left plot
of Figure \textbf{13} for different values of model parameter
$(\varsigma)$. The graph indicates positive and an asymmetrical
pattern in the behavior of the scale factor relative to time in the
exponential-bounce model II.
\begin{figure}
\epsfig{file=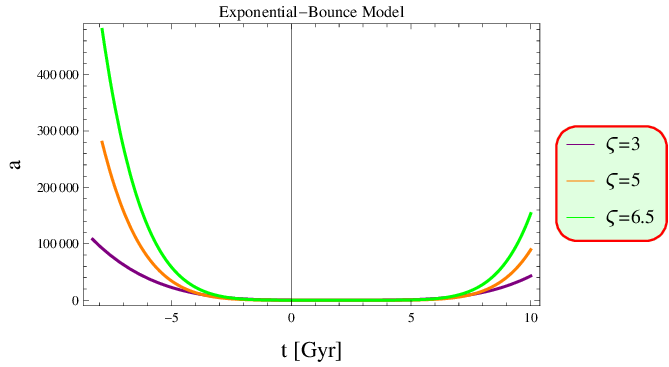,width=.5\linewidth}
\epsfig{file=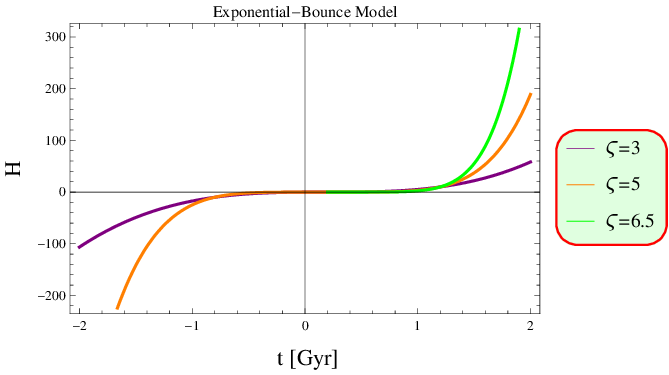,width=.5\linewidth}\caption{Behavior of scale
factor and Hubble parameter to cosmic time.}
\end{figure}

The corresponding Hubble parameter and non-metricity are given as
follows
\begin{equation}\label{30}
H=h_{0}(t-t_{b})^{\varsigma},~~Q=-6h_{0}^{2}(t-t_{b})^{2\varsigma}.
\end{equation}
The right plot of Figure \textbf{13} demonstrates that the Hubble
parameter becomes singular at bouncing point and this specific value
denotes the location of the bounce. Prior to the bounce, the Hubble
parameter is negative and becomes positive in the post-bounce phase.
This change signifies the transition from a contracting to an
expanding phases in the cosmic evolution. By applying Eq.(\ref{30})
in Eqs.(\ref{11b})-(\ref{11c}), we obtain
\begin{figure}
\epsfig{file=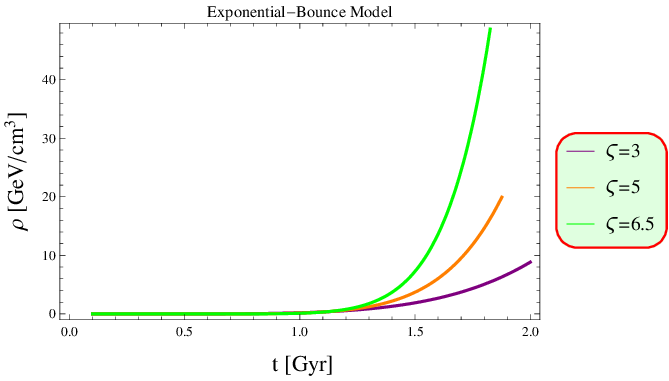,width=.5\linewidth}
\epsfig{file=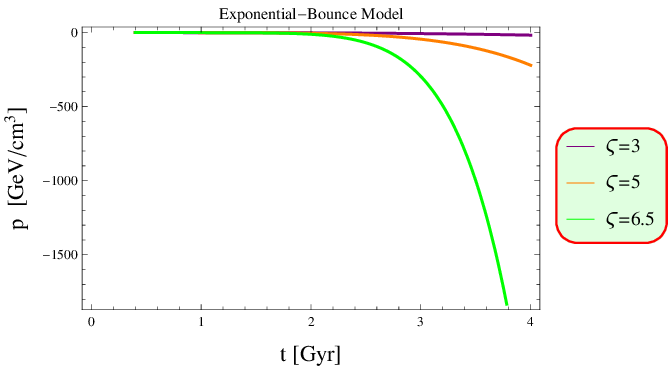,width=.5\linewidth}\caption{Behavior of energy
density and pressure for different values of $\varsigma$.}
\end{figure}
\begin{eqnarray}\label{31}
\rho&=&\frac{1}{(2\xi_{2}^{2}-49)}\big[h_{0}(t-t_{b})^{\varsigma-1}\xi_{1}
(-9h_{0}(t-t_{b})^{\varsigma-1})(\xi_{2}-7)+\varsigma\xi_{2}\big],
\\\nonumber
p&=&\frac{1}{(2\xi_{2}^{2}-49)}\big[h_{0}(t-t_{b})^{\varsigma-1}\xi_{1}
(9h_{0}(t-t_{b})^{\varsigma-1})(\xi_{2}+7)\\\label{32}
&+&\varsigma(3\xi_{2}+14)\big].
\end{eqnarray}
In Figure \textbf{14}, the energy density shows an upward trend
whereas the pressure demonstrates a downward pattern, aligning with
the anticipated behavior in the DE model. Using Eq.(\ref{31}) and
(\ref{32}), we get
\begin{equation}\label{32a}
\omega=\frac{(9h_{0}(t-t_{b})^{\varsigma+1})(\xi_{2}+7)
+\varsigma(3\xi_{2}+14)}{(-9h_{0}(t-t_{b})^{\varsigma+1})(\xi_{2}-7)+\varsigma\xi_{2})}
\end{equation}
The EoS parameter does not display symmetry around the bouncing
epoch and evolves in the phantom region as shown in Figure
\textbf{15}. In the context of $f(Q,T)$ gravity, this
exponential-bounce model II demonstrates behavior similar to the
exponential model I.
\begin{figure}\center
\epsfig{file=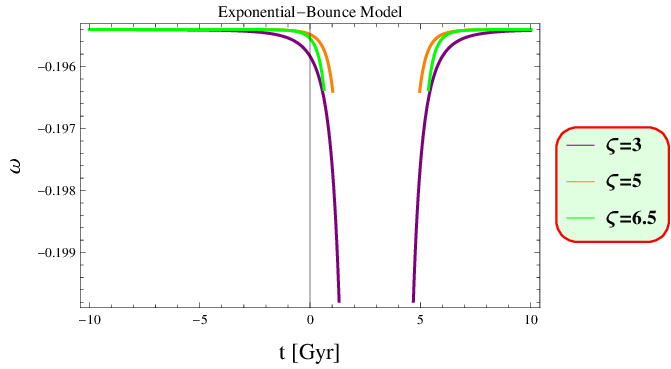,width=.5\linewidth}\caption{Behavior of EoS
parameter for different values of $\varsigma$.}
\end{figure}

\section{Analysis of Different Physical Aspects}

In this section, we explore a comprehensive analysis of various
physical aspects like deceleration parameter, energy conditions and
redshift analysis that influence the study of different cosmological
solutions. By examining the behavior of these parameters, we aim to
uncover insights that contribute to a deeper understanding of cosmic
dynamics.

\subsection{Deceleration Parameter}

The deceleration parameter $(q)$ is a dimensionless quantity which
measures the rate of expansion of the universe. It is defined as
\begin{equation}\label{33}
q=-\frac{a\dot{a}}{\dot{a}^{2}}=-1-\frac{\dot{H}}{{H}^{2}}.
\end{equation}
The positive value of deceleration parameter indicates an
decelerated cosmos whereas a negative value demonstrates an
accelerated universe. The asymmetrical nature of the deceleration
parameter is shown in the Figure \textbf{16}. The negative value of
deceleration parameter indicates that the universe is undergoing
accelerated expansion. This behavior aligns with observations of
distant supernova and the cosmic microwave background, which provide
evidence for the influence of dark energy driving the accelerated
expansion.
\begin{figure}
\epsfig{file=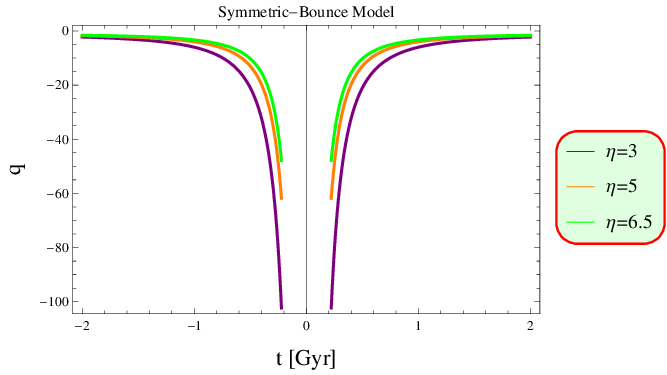,width=.5\linewidth}
\epsfig{file=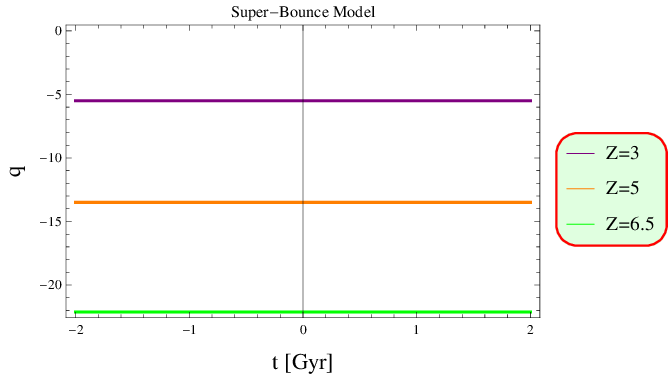,width=.5\linewidth}
\epsfig{file=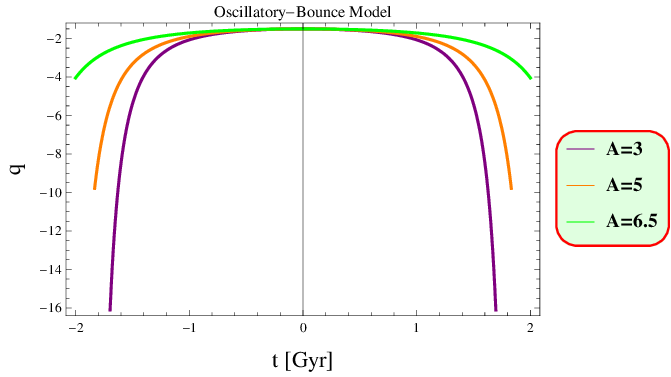,width=.5\linewidth}
\epsfig{file=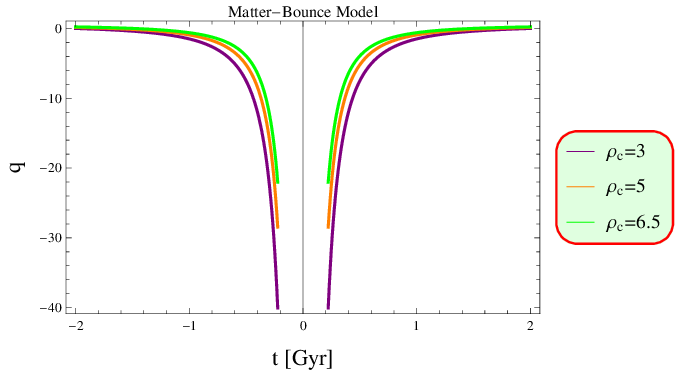,width=.5\linewidth}\center
\epsfig{file=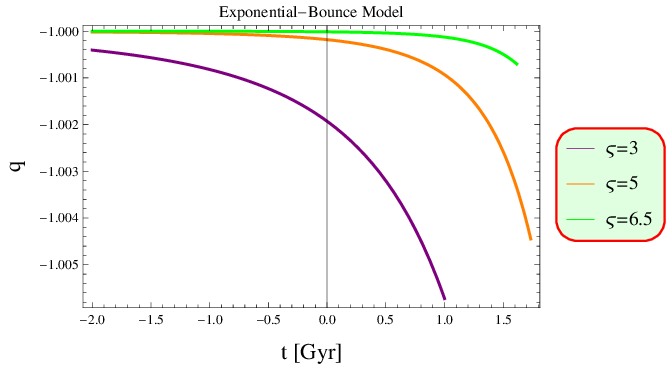,width=.5\linewidth}\caption{Behavior of
deceleration parameter for all the bouncing models.}
\end{figure}

\subsection{Analysis of Energy Conditions}
\begin{figure}
\epsfig{file=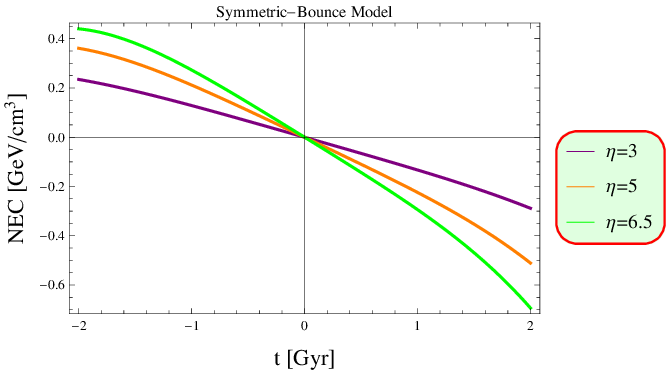,width=.5\linewidth}
\epsfig{file=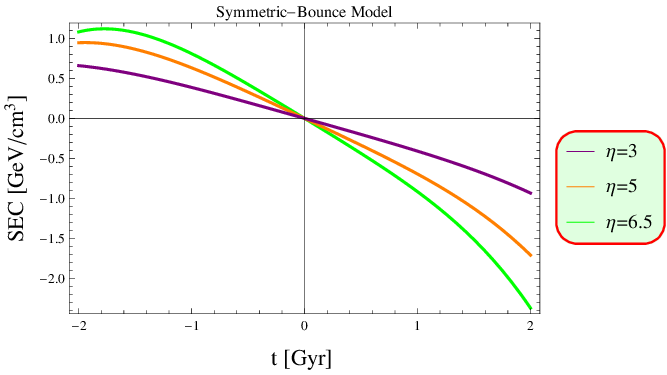,width=.5\linewidth}
\epsfig{file=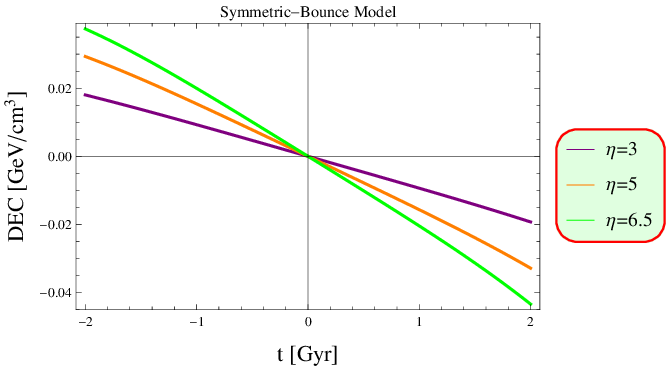,width=.5\linewidth}
\epsfig{file=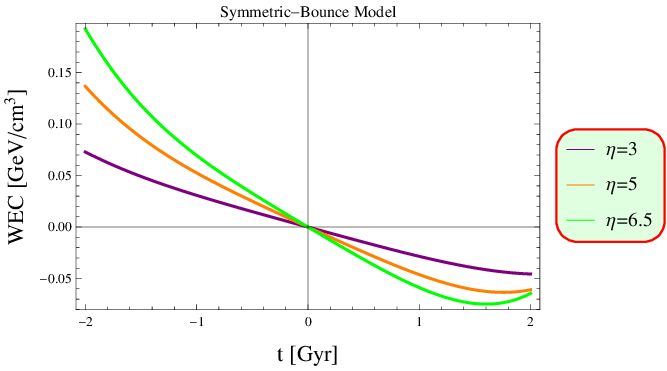,width=.5\linewidth}\caption{Behavior of energy
conditions for different parametric values.}
\end{figure}

Energy conditions are viable constraints with specific physical
properties based on the energy-momentum tensor used to assess the
physical consistency of cosmic models. Researchers impose these
constraints to evaluate the viability of different cosmic
configurations. The energy bounds are classified into several types
as null energy condition ($0\leq\rho+p$), dominant energy condition
($0\leq\rho$, $0\leq\rho+p$), weak energy condition ($0\leq\rho$, $0
\leq\rho\pm p$) and strong energy condition ($0\leq\rho+p$, $0\leq
\rho+3p$). In this study, we present a graphical representation of
these energy constraints for all considered bouncing cosmological
models. By examining these conditions, we can understand the
characteristics of cosmic geometries and their relationship to the
EMT. The violation of the null energy condition implies the
violation of all other energy conditions, which guarantees the
existence of a non-singular bounce \cite{1z10a}.
\begin{figure}
\epsfig{file=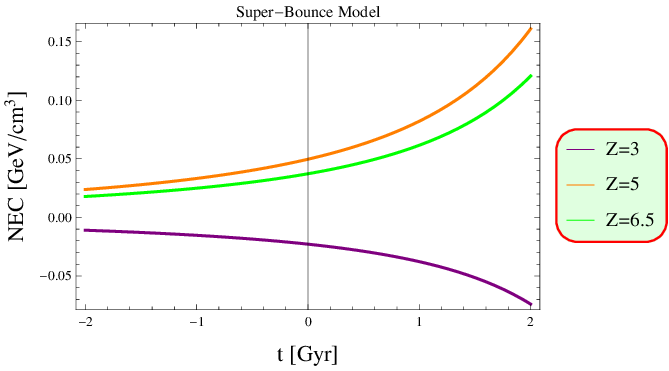,width=.5\linewidth}
\epsfig{file=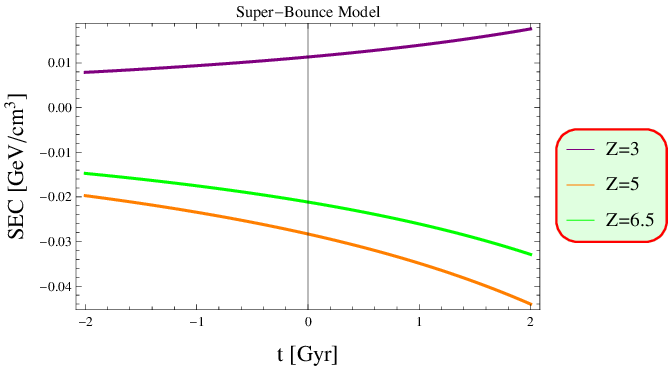,width=.5\linewidth}
\epsfig{file=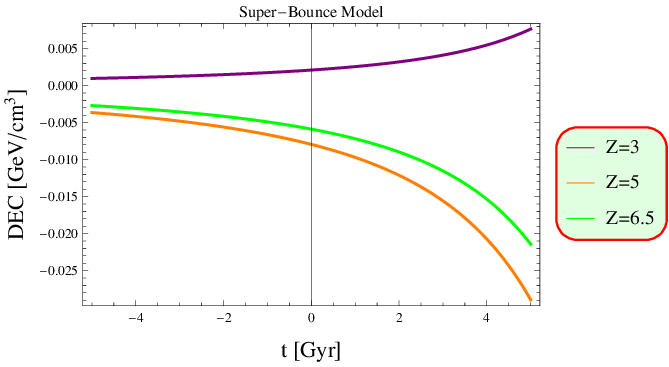,width=.5\linewidth}
\epsfig{file=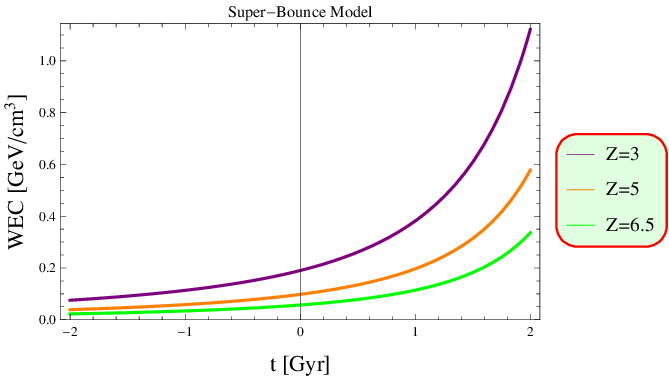,width=.5\linewidth}\caption{Analysis of energy
conditions corresponding to cosmic time.}
\end{figure}
\begin{figure}
\epsfig{file=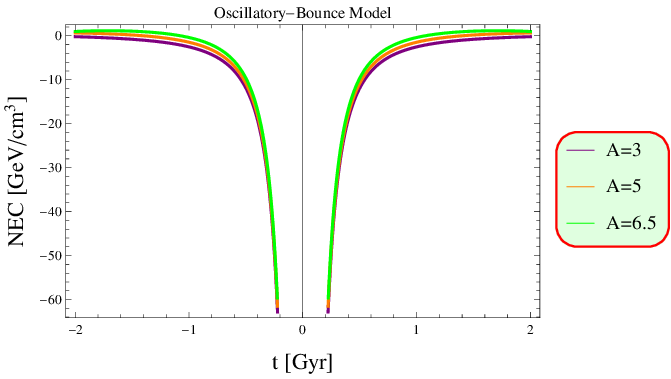,width=.5\linewidth}
\epsfig{file=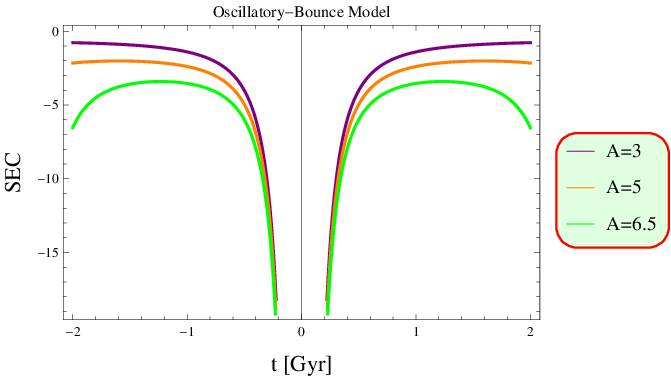,width=.5\linewidth}
\epsfig{file=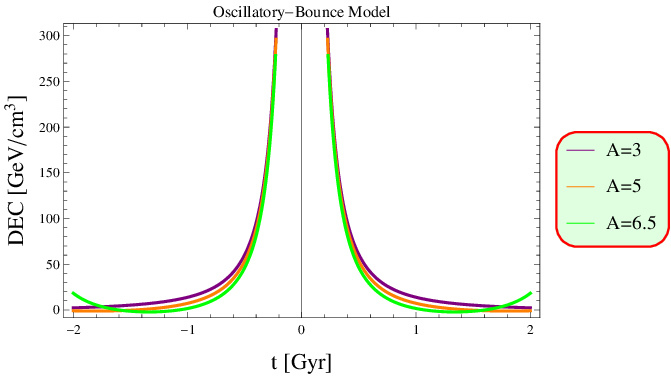,width=.5\linewidth}
\epsfig{file=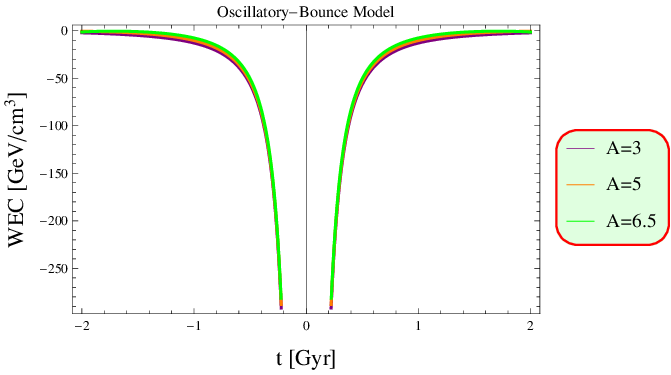,width=.5\linewidth}\caption{Graphs of energy
constraints for oscillatory-bounce model.}
\end{figure}

Figures \textbf{17-21} demonstrate that the bouncing criteria of the
cosmos is satisfied for all the considered models, providing a
comprehensive analysis of these conditions under which the universe
exhibit a bounce epoch, ensuring the avoidance of singularities. The
detailed graphical representation supports the stability and
consistency of non-singular bounce in this theoretical framework.
These results underscore the interplay between energy condition
violations and the bouncing behavior. The violation of energy
conditions is a critical requirement for the realization of a
non-singular bounce.
\begin{figure}
\epsfig{file=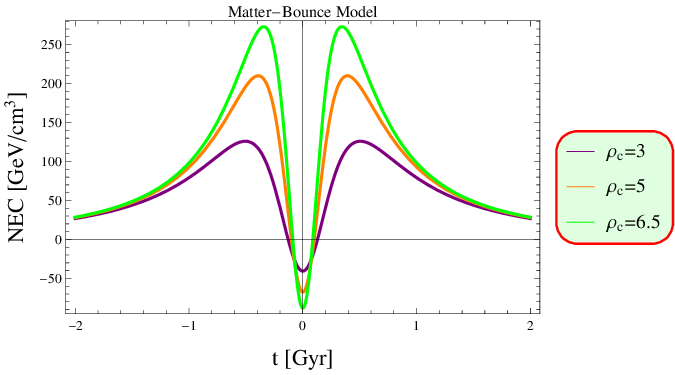,width=.5\linewidth}
\epsfig{file=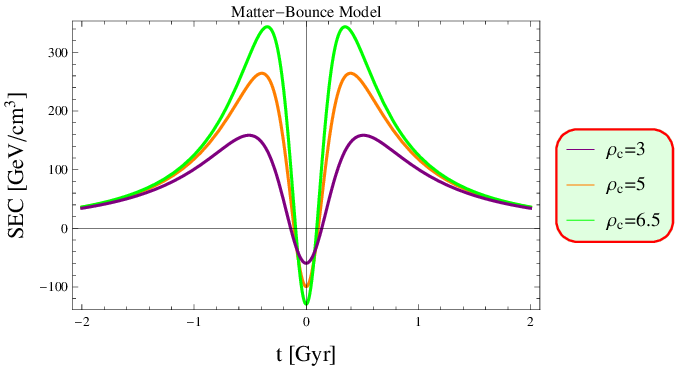,width=.5\linewidth}
\epsfig{file=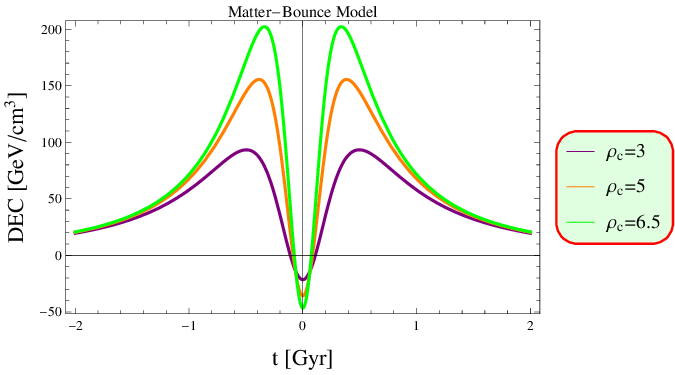,width=.5\linewidth}
\epsfig{file=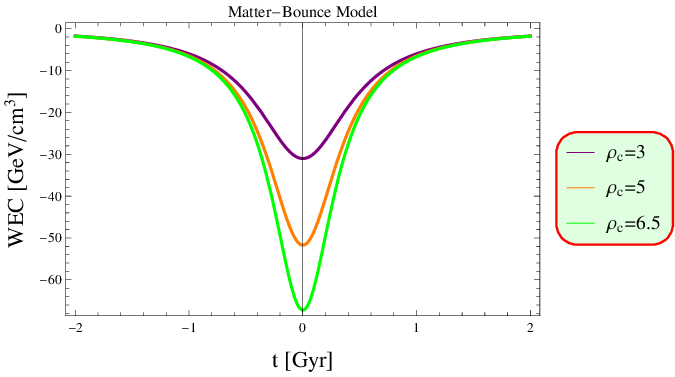,width=.5\linewidth}\caption{Evaluation of
energy conditions for different values of $\rho_{c}$.}
\end{figure}
\begin{figure}
\epsfig{file=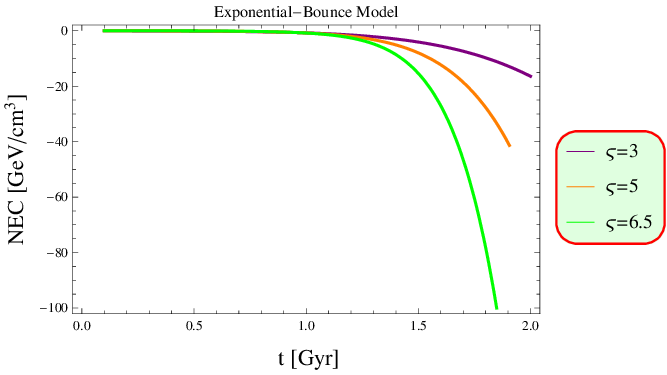,width=.5\linewidth}
\epsfig{file=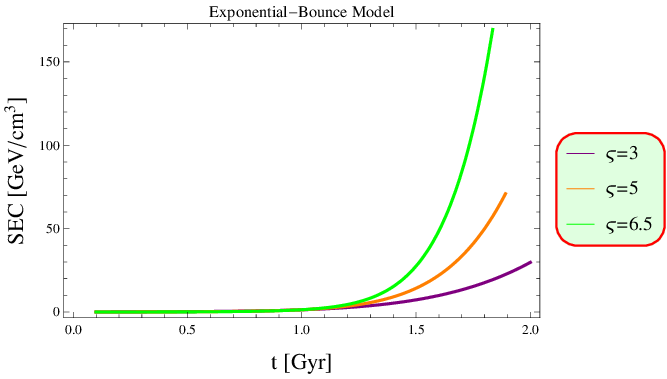,width=.5\linewidth}
\epsfig{file=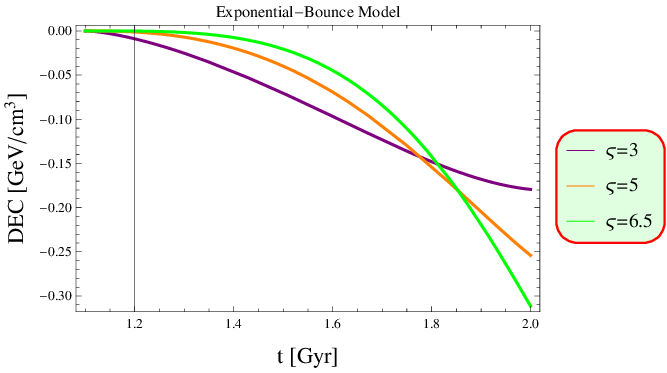,width=.5\linewidth}
\epsfig{file=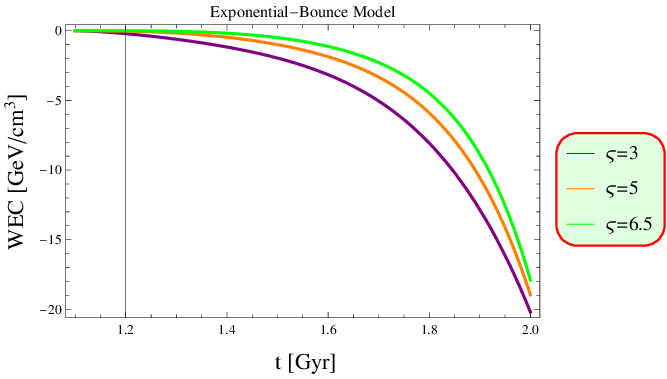,width=.5\linewidth}\caption{Behavior of energy
bounds for exponential-bounce model.}
\end{figure}

\subsection{Evaluation of Redshift}

In this section, we examine the redshift parameter $(z)$ to study
the behavior of matter configuration and cosmological scenario. The
scale factor is given by $a(t)=a_{0}t^{\varphi}$, where $\varphi$ is
an arbitrary constant and we assume the current value of $a_{0}$ as
1. The value of deceleration parameter is used as defined in
Eq.(\ref{33}). By using the value of $\varphi=\frac{1}{1+q}$ in
scale factor, we have
\begin{equation}\label{34}
a(t)=t^{\frac{1}{1+q}},
\end{equation}
where $q=-0.831^{+0.091}_{-0.091}$ \cite{100}. The rate at which the
universe is currently expanding can be described as
\begin{equation}\label{35}
H=\frac{\dot{a}}{a}=H=(1+q)^{-1} t^{-1}, \quad  H_{0}=(1+q)^{-1}
t_{0}^{-1}.
\end{equation}
The expansion of the universe is affected by parameters such as $q$
and $H_{0}$. To analyze the connection between the redshift
parameter and the scale factor, we have
\begin{equation}\label{36}
H=H_{0}(1+z)^{1+q}, \quad \dot{H}=-H_{0}(1+z)^{2+2q}.
\end{equation}
The value of non-metricity in Eq.(\ref{11}) is determined as
\begin{equation}\label{37}
Q=-6H_{0}^{2}(1+z)^{2+2q}.
\end{equation}
The field equations in terms of redshift are given by
\begin{eqnarray}\label{38}
\rho&=&\frac{-1}{49+2(\xi_{2}-7)\xi_{2}}[9\xi_{1}(\xi_{2}-7)H_{0}^{2}
(z+1)^{2q+2}],
\\\label{39}
p&=&\frac{1}{(49+2(\xi_{2}^{2}-7)\xi_{2}}[9\xi_{1}(\xi_{2}-7)H_{0}^{2}
(z+1)^{2q+2}].
\end{eqnarray}
\begin{figure}
\epsfig{file=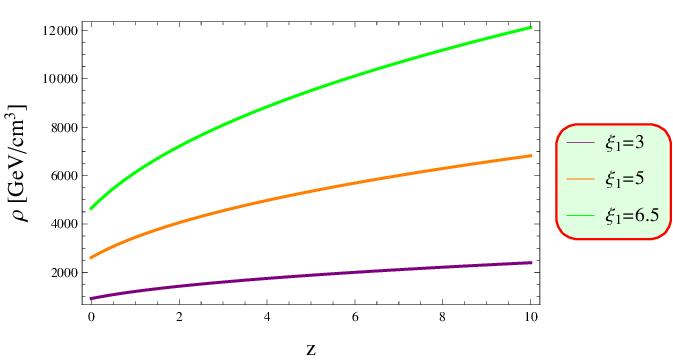,width=.5\linewidth}
\epsfig{file=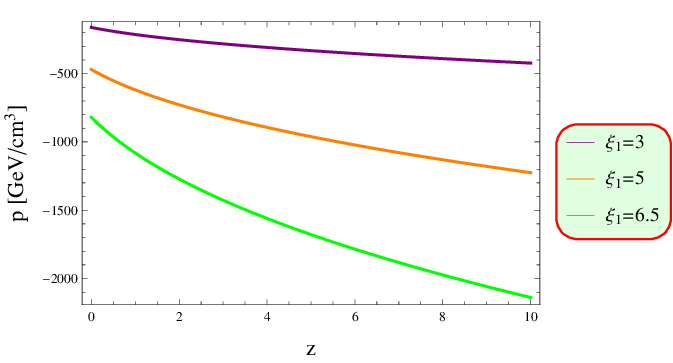,width=.5\linewidth}\caption{Behavior of matter
variables corresponding to redshift.}
\end{figure}
\begin{figure}\center
\epsfig{file=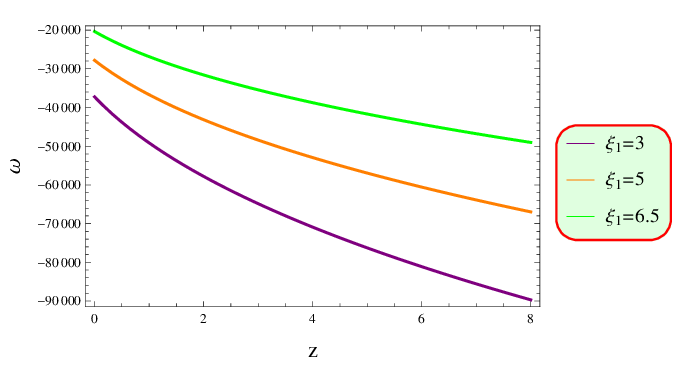,width=.5\linewidth}\caption{Behavior of EoS
parameter versus redshift.}
\end{figure}

Figure \textbf{22} shows the variation of energy density and
pressure as function of redshift for the considered $f(Q,T)$ model.
The graphical behavior shows that the energy density increases and
remains positive while the pressure decreases negatively which is
consistent with the behavior of the DE. The EoS parameter is
essential in cosmology because it describes the relationship between
the pressure and density of the cosmic matter. Figure \textbf{23}
demonstrates the graphical behavior of the EoS parameter as a
function of redshift. This parameter shows the phantom region
$(\omega < -1)$ which indicates an accelerated expansion of the
universe.

\section{Final Remarks}

Recent cosmological observations, including measurements of cosmic
microwave background radiation, Planck data, supernovae type-Ia,
large-scale structures and galaxy redshift surveys indicate that the
expansion of the cosmos is accelerating \cite{1z11}-\cite{1z14}.
This discovery has led to the paradigm of modified theories of
gravity as a fundamental framework for understanding gravitational
interactions and their influence on cosmic expansion. These modified
theories incorporate additional gravitational fields, spatial
dimensions and higher-order derivatives offering diverse approaches
to explain the phenomenon of cosmic acceleration through
modifications to EGTR. The big bang cosmology faces significant
challenges with the initial singularity and inflationary paradigm
prompting diverse solutions in the literature. Several methods have
been employed in the literature to address this issue and bouncing
cosmology is considered as one of the most effective alternatives.
Additionally, modified gravity offers a promising framework for
developing new cosmological models that can eliminate the
long-standing cosmological challenges. This study aims to
investigate the singularity issue in the context of $f(Q,T)$ gravity
using bouncing cosmology.

The motivation to investigate the bouncing cosmology in the
framework of $f(Q,T)$ theory is to examine the viability of
non-singular bouncing solutions. By using different constraints, it
is possible to assess the validity of this theory in comparison to
other cosmic models. Therefore, integrating bouncing cosmology in
the modified $f(Q,T)$ theory offers a platform to address the
fundamental inquiries in cosmology such as the cosmic origin,
DE/dark matter nature and behavior of gravity at classic and quantum
levels. The main goal of this research is to examine how $f(Q,T)$
gravity contributes to construct effective cosmological models which
explore the issue of accelerated expansion of the cosmos and its
potential implications. The rate of cosmic expansion is determined
by using different scale factors and the Hubble parameters. There
are two scenarios to consider , i.e., the scale factor approaches to
zero (big bang singularity) or  considering the bouncing models. In
the cosmic bounce models, the scale factor never reaches to zero,
thus avoiding any spacetime singularity. Initially, the cosmos
expands, then contracts and this cycle of expansion and contraction
continues until to reach a minimum size of the scale factor.

We have focused on investigating the familiar bouncing cosmological
scenarios in a flat FRW spacetime characterized by a perfect fluid
matter distribution. Our study encompasses five distinct types of
bouncing solutions, including symmetric-bounce, super-bounce,
oscillatory-bounce, matter-bounce, and exponential bouncing models.
We have analyzed the cosmological parameters including scale factor,
Hubble parameter, EoS parameter, deceleration parameter and behavior
of energy conditions associated with each of these solutions. The
main findings are summarized as follows.
\begin{itemize}
\item
The scale factor is a positive function that varies over time to
describe the changing size of the cosmos, reflecting its expansion
dynamics with respect to cosmic time (left plot of Figures
\textbf{1}, \textbf{4}, \textbf{7}, \textbf{10} and \textbf{13}).
These graphs exhibit a symmetrical pattern illustrating that these
scale factors increase and decrease evenly on both sides of a
bouncing point except exponential model II scale factor.
\item
The Hubble parameter is negative during the pre-bounce phase
indicating a contracting cosmos. As the universe approaches the
critical bouncing point $(t=0)$, the Hubble parameter approaches to
zero. Moving into the post-bounce epoch, the Hubble parameter
becomes positive which signifies the cosmic expansion. This
progression through the phases describes the dynamic nature of all
the considered cosmic model as it transitions between contraction
and expansion phases.
\item
The graphical behavior of energy density and pressure reveals a an
inverse relationship between energy density and pressure which is in
accordance with the expected behavior predicted by the DE model.
\item
The EoS parameter helps us to comprehend how different energy
components interact and influence the overall dynamics during the
bouncing point. At the point of bounce, the EoS parameter becomes
singular and undergoes rapid change. It shows symmetry around the
bounce epoch and transitions into the phantom region. This change
signifies a significant shift in the characteristics of this
parameter as it moves towards the bouncing point.
\item
The behavior of deceleration parameter is crucial to understand the
cosmic dynamics. The deceleration parameter shows the cosmic
accelerated expansion as shown in Figure \textbf{16}. This behavior
is consistent with the cosmological observations and also describes
the role of dark energy in driving this accelerated expansion.
\item
We have discussed the cosmic acceleration and expansion through
energy conditions in this framework. The NEC is violated for all the
bouncing models which ensures the existence of non-singular cosmic
bounce in this modified framework (Figures \textbf{17-21}).
\item
The energy density in terms of redshift function shows positively
increasing behavior while pressure demonstrates the negatively
decreasing behavior for the considered $f(Q,T)$ model, confirming
that the cosmos is in the expansion phase (Figure \textbf{22}).
\item
The graphical behavior of the EoS parameter corresponding to
redshift represents the phantom region, which shows the cosmic
accelerated expansion (Figure \textbf{23}).
\end{itemize}

We have examined the existence of non-singular cosmic bounce models
under the influence of modified $f(Q,T)$ terms. Our analysis on the
physical aspects has unveiled a feasible profile of cosmological
models. We have investigated several fundamental bounce models
including symmetric-bounce, super-bounce, oscillatory-bounce,
matter-bounce and exponential-bounce that delineate the cosmic
evolution. It is worth noting that all these models indicate the
presence of viable bouncing cosmology in this theoretical framework.
In a recent paper, Gadbail et al \cite{1z15} investigated various
cosmic bounce models in $f(Q)$ gravity and explored cosmic evolution
using fixed parametric values. Agrawal and his colleagues
\cite{1z16} studied the matter-bounce scenario in $f(R,T)$ theory
for various parametric values greater than 1. We have done graphical
analysis for both values less and greater than 1. Our findings not
only coincide with the existing literature but also provide further
understanding of the dynamics of cosmic evolution. Our results
demonstrate that cosmic solutions remain feasible even when
considering modified terms, highlighting the
strength of our theoretical framework. \\\\
\textbf{Data Availability:} No data was used for the research
described in this paper.

\end{document}